\documentclass[11pt,a4paper]{article}
\usepackage{jheppub}
\usepackage{graphicx}
\usepackage{amsmath}
\usepackage{bm}
\def\Vec#1{\mbox{\boldmath $#1$}}

\def\itmb{\begin{itemize}}
\def\itme{\end{itemize}}
\def\enmb{\begin{enumerate}}
\def\enme{\end{enumerate}}
\def\eqnb{\begin{equation}}
\def\eqne{\end{equation}}

\def\PLB{{Phys. Lett.} B}
\def\PRL{Phys. Rev. Lett.}
\def\PRD{{Phys. Rev.} D}

\title{Cartan's Supersymmetry and the Decay of a $h^0$\\
 with the mass $m_{h^0}\simeq 11$GeV
 to $\Upsilon(nS)\gamma (n=1,2,3)$ }  
\author{Sadataka Furui}
\affiliation{
 Graduate School of Teikyo University\\
2-17-12 Toyosatodai, Utsunomiya, 320-0003 Japan}
\emailAdd{ furui@umb.teikyo-u.ac.jp}

\abstract{
In the LHCb detector at CERN, decays of $\chi_b(3P)$ meson to $\Upsilon(1S)\gamma$ and $\Upsilon(2S)\gamma$ are reported at centre-of-mass energy of $\sqrt s=7$ and 8 TeV.
Following the success of the assignment of $\chi_b(1P)\to\Upsilon(1S)\gamma$ of the mass of $m(\chi_b(1P))=9.8923$ GeV and $\chi_b(2P)\to\Upsilon(1S)\gamma$ of the mass of $m(\chi_b(2P))=10.2547$ GeV, the new state $\chi_b(3P)$ of the mass of 10.5157 GeV was assigned, but its $J^{P}$ was not fixed.

We study the possibility that this boson is the light Higgs boson $h^0(0^+)$, and study its decay modes to a $b\bar b$ which reduces to an $\Upsilon(mS)$ ($m=1,2$ or 3) and $\bar q q$ which reduces to a $\gamma$, using the Cartan's supersymmetry. Recent non observation of polarizations of $\Upsilon(nS)$ by the CMS collaboration is consistent with the theory. }
\begin{document}
\maketitle

\section{Introduction}
In 2012, the ATLAS Collaboration presented a new $\chi_b$ state in radiative transition to $\Upsilon(1S)$ and $\Upsilon(2S)$\cite{ATLAS12b,Abazov12}. Since radiative transitions of $\chi_b(1P)$ and $\chi_b(2P)$ were observed near the energy of the new state, the state was assigned as $\chi_b(3P)$, but its $J^P$ could not be fixed by experimental analyses\cite{CMS12,LHCb14}.
Recently the CMS collaboration found $\Upsilon(3S)$ although backgrounds are large, and that polarizations of $\Upsilon(nS)$ (n=1,2,3) are very small \cite{CMS16}.

The smallness of the polarization suggests that the parent $\chi_b(nP)$ states have $J=0$.
We propose an assignment of the new state as the partner of the Higgs boson $H^0$ discovered by the ATLAS collaboration\cite{ATLAS12a}, which has a lower mass and is called $h^0$\cite{Labelle10}. 
Higgs field $H$ belongs to $SU(2)_L$ doublet and can be expressed as
\[
H_u=\left(\begin{array}{c} H_u^+\\
                                H_u^0\end{array}\right), \quad
H_d=\left(\begin{array}{c} H_d^0\\
                                H_d^-\end{array}\right), 
\]

The masses of the gauge bosons are
\begin{eqnarray}
{\mathcal L}_{MGB}&=&\frac{\nu^2}{4}g^2(W_1^2+W_2^2)+\frac{\nu^2}{4}(gW_3-g' B)^2\nonumber\\
&=&m_W^2 W^{+\mu}{W^-}_\mu+\frac{1}{2}m_Z^2 Z_\mu Z^\mu,
\end{eqnarray}
where minimum of the potential is 
\[
\langle H_u\rangle_{min}=\left(\begin{array}{c}0\\
                                                                   \nu_u\end{array}\right),\qquad 
\langle H_d\rangle_{min}=\left(\begin{array}{c}\nu_d\\
                                                                   0\end{array}\right)
\]
and $\nu^2=\nu_u^2+\nu_d^2$.

The possible masses of the Higgs particles are
\[
m_{H^\pm}^2=m_W^2+m_{A^0}^2
\]
for charged massive states, and
\begin{eqnarray}
m_{h^0}^2&=&\frac{m_{A^0}^2+m_Z^2}{2}-\frac{1}{2}\sqrt{(m_{A^0}^2+m_Z^2)^2-4m_{A^0}^2m_Z^2\cos^2 2\beta}\nonumber\\
m_{H^0}^2&=&\frac{m_{A^0}^2+m_Z^2}{2}+\frac{1}{2}\sqrt{(m_{A^0}^2+m_Z^2)^2-4m_{A^0}^2m_Z^2\cos^2 2\beta},
\end{eqnarray}
where $\tan\beta=\nu_u/\nu_d$, for neutral massive states.

When $\cos 2\beta=0$, $m_{h^0}^2=0$, $m_{H^0}^2=m_{A^0}^2+m_Z^2$, and $m_Z=91.2$GeV, $m_{H^0}=125$GeV\cite{ATLAS12a} gives 
\[
m_{A^0}=85.5 \rm{GeV}.
\]
and $m_W=80.4$ GeV yields $m_{H^\pm}=117$ GeV.

There is a report of the search of $H^+$\cite{CMS12} using the $t\to H^+ b$ decay and $H^+\to \tau \nu_\tau$ which yields $m_{H^+}=120$GeV, but in this analysis, the branching fraction $B(H^+\to\tau\nu_\tau)$ could not be well determind, and it was assumed to be equal to 1. We expect that it is due to instability of the $H^+$ state. The requirement that $m_{H^\pm}^2=m_W^2+m_{A^0}^2=(120$ GeV)$^2$ gives $m_{A^0}=78.0$ GeV and $m_{H^0}$ becomes 125 GeV, with 
\[
m_{A^0}=78.0 \rm{GeV \quad and} \quad \cos 2\beta=\pm 0.1878.
\]

The fixed $\cos 2\beta$ gives the mass squared of $h^0$ 
\[
m_{h^0}^2=\frac{m_{A^0}^2+{m_Z}^2}{2}-\frac{1}{2}\sqrt{(m_{A^0}^2+{m_Z}^2)^2-4{m_{A^0}}^2{m_Z}^2\cos^2 2\beta}=(11.2{\rm GeV})^2.
\]

Near this energy region there are $\chi_{b0}(1P, J^{PC}=0^{++}, 9.86$ GeV), $\chi_{b1}(1P, J^{PC}=1^{++}, 9.89$ GeV) and  $\chi_{b2}(2P, J^{PC}=2^{++}, 10.23$ GeV) which are expected to be made of $b\bar b$ and a state which is called $\chi_b(3P , 10.53$ GeV).  The scalar boson $\chi_b(3P)$ decays radiatively to $\Upsilon$(1S) and $\Upsilon$(2S), and its $C=+$ but its $J^P$ is not well known\cite{ATLAS12b,Abazov12,PDG14}.

 The mass of $\chi_b(3P)$ is slightly below the $B\bar B$ threshold and there remains a possibility that the SUSY-breaking potential\cite{Labelle10},
\begin{eqnarray}
V_{SSB}&=&v (H_u^+  H_u^0) i\tau_2 \left(\begin{array}{c} H_d^0\\
                                                                                 H_d^-\end{array}\right)
            +v^*(H_d^{0\dagger} H_d^{-\dagger})(- i\tau_2)\left(\begin{array}{c} H_u^{+\dagger}\\
                                                              H_u^{0\dagger}\end{array}\right)\nonumber\\
         &=&v(H_u^+ H_d^- -H_u^0 H_d^0) +h.c.
\end{eqnarray}
where $v>0$, makes a $H^\pm$ unstable, and a $h^0$ appears as a $\chi_b(3P)$.  

\section{$h^0(0^+)\to \Upsilon(mS) q \bar q \to \Upsilon(mS)\gamma$}
In the standard supersymmetric theory, coupling of lepton pairs $\ell\bar\ell$ and $\Upsilon(b\bar b)$ 
occur through two Higgs bosons \cite{CMS12}, as $\ell\bar\ell H\bar H b\bar b$. Cartan's supersymmetry has different character from standard supersymmetric theory. It contains transformations of vector fields and leptons inside and it is not necessary to introduce gluinos et al. 

In order to study the decay of $h^0(0^+)$ into $\Upsilon(mS)\gamma$, where $m=1,2$ or 3, I adopt the model similar to that used in $H^0\to\ell\bar\ell\ell\bar\ell\to 2\gamma$ \cite{SF12a,SF12b,SF12c,SF13a,SF13b}. 
In the present case I consider Higgs field defined by a complex number $x_i+i\, x_j$ ($i\ne j, 1,2,3)$ produces antiquarks $\xi_{i4}$ and $\xi_{j4}$ or quarks $\xi_i$ and $\xi_j$. Produced antiquarks $\xi_{i4}$ interacts with vector fields $X'$ and produces $\ell(b)$ in $\Upsilon$, and the produced antiquark $\xi_{j4}$ interacts with other vector field $X'$ and produces $\ell$ which are combined with $\bar \ell(\bar b)$ in the final states become photons and transform to lepton-antilepton pairs.  
The $b$ and the $\bar b$ becomes correlated to make $\Upsilon(1S,J^{PC}=1^{--},9.46$GeV),  $\Upsilon(2S,J^{PC}=1^{--},10.02$GeV) and $\Upsilon(3S,J^{PC}=1^{--},10.35$GeV) . 

Typical diagrams of $\Upsilon_1$in which helicity of the $b\bar b$ are chosen along $\Vec j$ or $\Vec k$ are shown in Figures 1, those of $\Upsilon_2$in which helicity of the $b\bar b$ are chosen along $\Vec i$ or $\Vec k$ are shown in Figures 2, and those of $\Upsilon_3$in which helicity of the $b\bar b$ are chosen along $\Vec i$ or $\Vec j$ are shown in Figures 3. The spontaneous symmetry breaking in the standard model contains the field $W_1\pm i\,W_2$ which corresponds to our $x_1+i\,x_2$ and a comination of $W_3$ and $B$ which corresponds to our $x_3$. In our case, symmetrybreakings of $\Upsilon(1S)$, $\Upsilon(2S)$ and $\Upsilon(3S)$ are relatively small, and $x_1,x_2$ and $x_3$ are assumed to be symmetric, consistent with small polarization \cite{CMS16}. 
In these diagrams, stable points of Higgs potential are denoted by $h_0(x_i+i\,x_j)$ for quarks, and $h_0'(x_i+i\,x_j)$ for antiquarks. 

We first define the helicity of the produced quark on the upper circle $\psi$ or $C\psi$ in the case of the decay of $h_0$, and $\phi$ or $C \phi$ in the case of the decay of ${h_0}'$. Production of $\bar b$ or $b$ quark by emission of a vector particle $x0$ or $x2$ is assigned using the rule of ${^t \phi}CX \psi$ of Cartan\cite{SF12a}, which is defined as
\begin{eqnarray}
{\mathcal F}&=&^t\phi{\mathcal C}X\psi={^t\phi} \gamma_0x^\mu\gamma_\mu\psi\nonumber\\
&=&x^1(\xi_{12}\xi_{314}-\xi_{31}\xi_{124}-\xi_{14}\xi_{123}+\xi_{1234}\xi_1)\nonumber\\
&+&x^2(\xi_{23}\xi_{124}-\xi_{12}\xi_{234}-\xi_{24}\xi_{123}+\xi_{1234}\xi_2)\nonumber\\
&+&x^3(\xi_{31}\xi_{234}-\xi_{23}\xi_{314}-\xi_{34}\xi_{123}+\xi_{1234}\xi_3)\nonumber\\
&+&x^4(-\xi_{14}\xi_{234}-\xi_{24}\xi_{314}-\xi_{34}\xi_{124}+\xi_{1234}\xi_4)\nonumber\\
&+&x^{1'}(-\xi_{0}\xi_{234}+\xi_{23}\xi_{4}-\xi_{24}\xi_{3}+\xi_{34}\xi_2)\nonumber\\
&+&x^{2'}(-\xi_{0}\xi_{314}+\xi_{31}\xi_{4}-\xi_{34}\xi_{1}+\xi_{14}\xi_3)\nonumber\\
&+&x^{3'}(-\xi_{0}\xi_{124}+\xi_{12}\xi_{4}-\xi_{14}\xi_{2}+\xi_{24}\xi_1)\nonumber\\
&+&x^{4'}(\xi_{0}\xi_{123}-\xi_{23}\xi_{1}-\xi_{31}\xi_{2}-\xi_{12}\xi_3).
\end{eqnarray}

Cartan's trilinear form of the lepton antilepton and vector field coupling allows coupling of $\Upsilon_i$ (i=1,2,3) fields that couples to $b\bar b$ which are expressed by $\ell\bar\ell$ in the left upper corner  and $q\bar q$ in the bottom.  The $q\bar q$ decays to photons and couple to $\mu\bar\mu$.  In \cite{SF16}, I considered Higgs boson coupling to $W^\pm$ and $Z$ bosons, by assigning complex fields $x+i\,y$ in the vacuum. The value $x^2+y^2$ is related to the value of the Higgs potential. 

\begin{figure}[htb]
\begin{minipage}[b]{0.47\linewidth}
\begin{center}
\includegraphics[width=4cm,angle=0,clip]{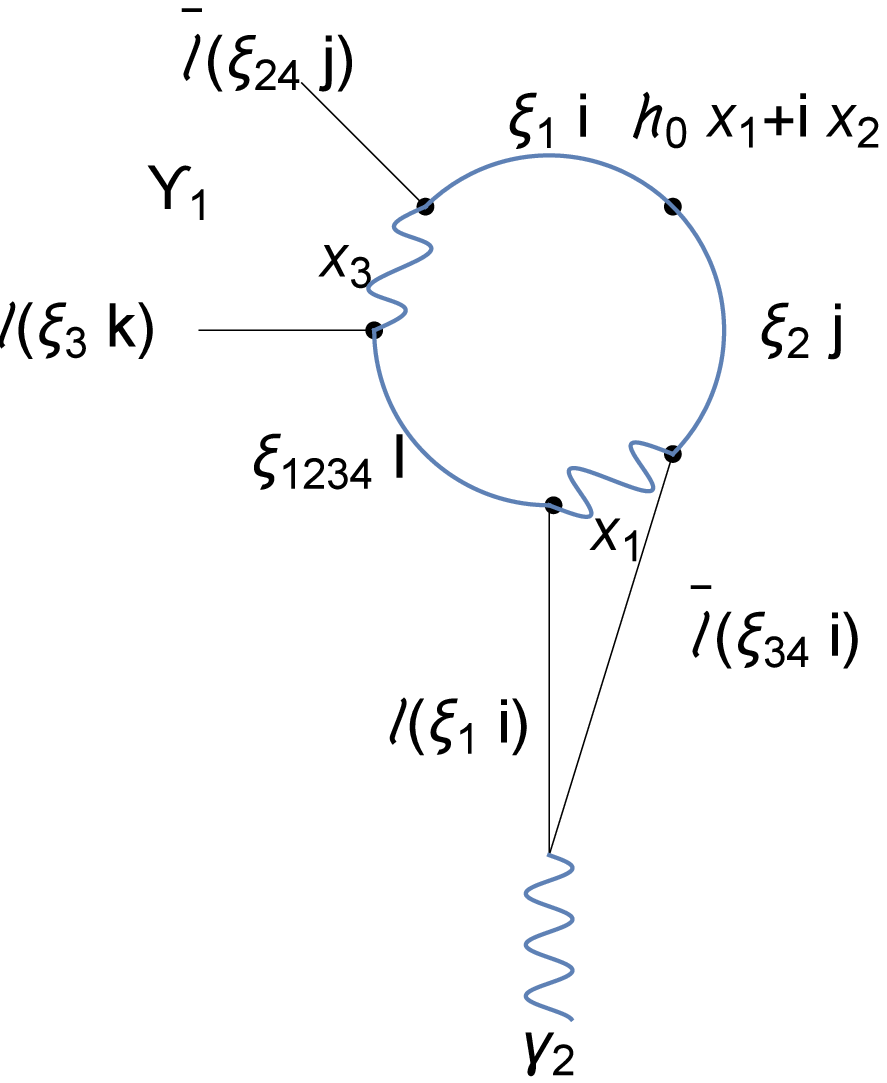}   
\end{center}
\end{minipage}
\hfill
\begin{minipage}[b]{0.47\linewidth}
\begin{center}
\includegraphics[width=4cm,angle=0,clip]{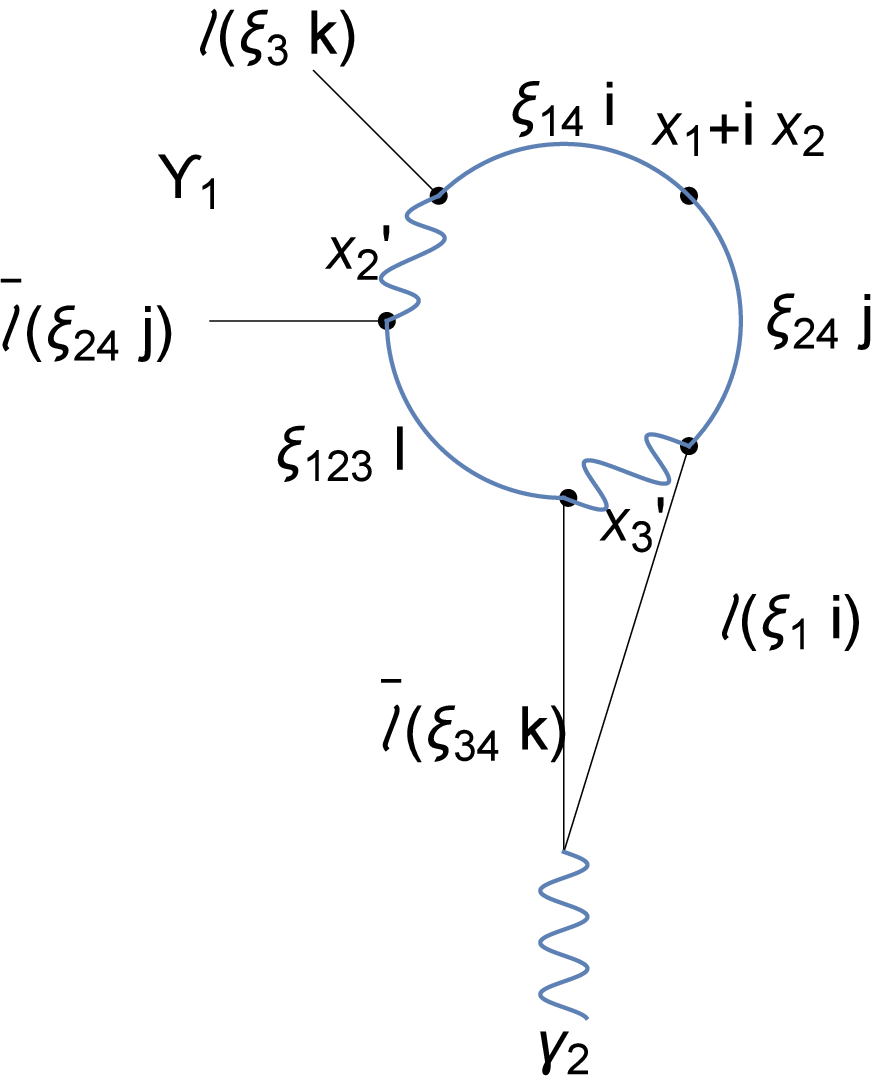}  
\end{center}
\end{minipage}
\begin{minipage}[b]{0.47\linewidth}
\begin{center}
\includegraphics[width=4cm,angle=0,clip]{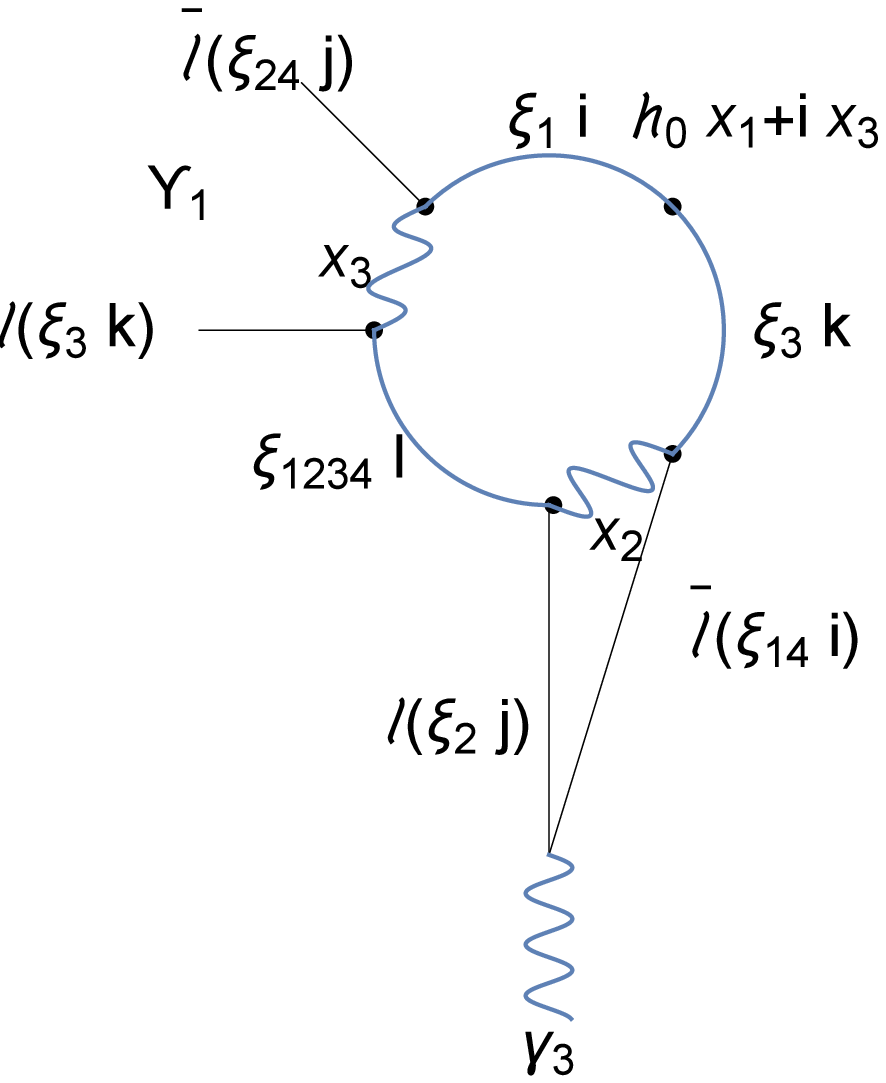} 
\end{center}
\end{minipage}
\hfill
\begin{minipage}[b]{0.47\linewidth}
\begin{center}
\includegraphics[width=4cm,angle=0,clip]{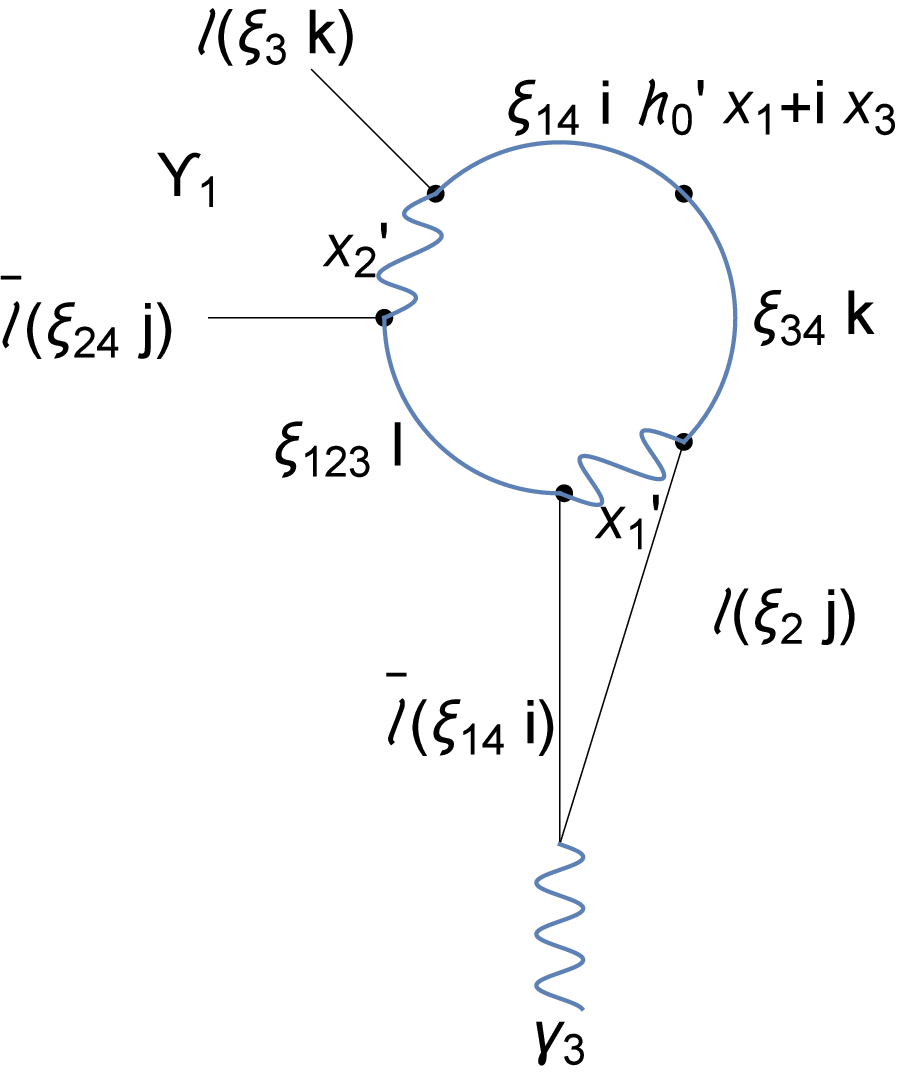}  
\end{center}
\end{minipage}
\caption{The $h(0^+, X+i\,Y)\to \Upsilon_1(j\times k)\gamma_j$ (j=2,3) decay. $\Upsilon_1$ is described by $\ell\bar\ell$ without interactions. }
\end{figure}
\begin{figure}[htb]
\begin{minipage}[b]{0.47\linewidth}
\begin{center}
\includegraphics[width=4cm,angle=0,clip]{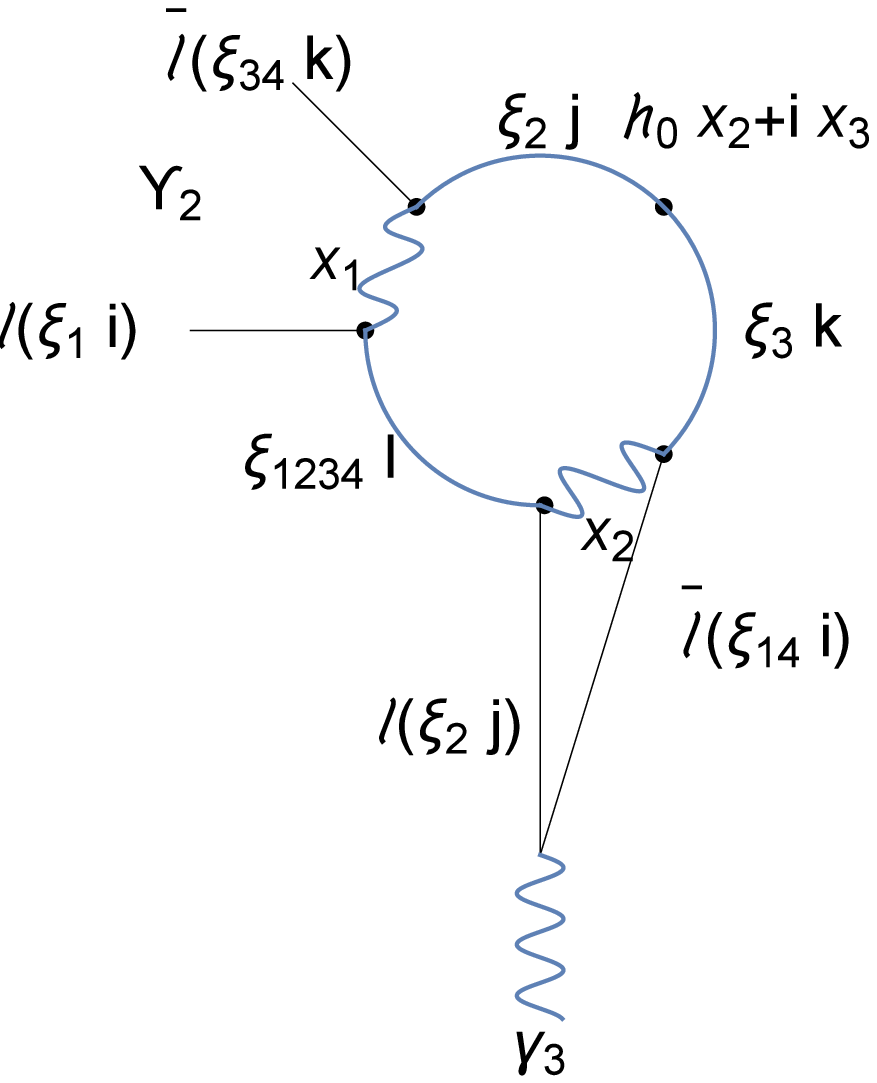}   
\end{center}
\end{minipage}
\hfill
\begin{minipage}[b]{0.47\linewidth}
\begin{center}
\includegraphics[width=4cm,angle=0,clip]{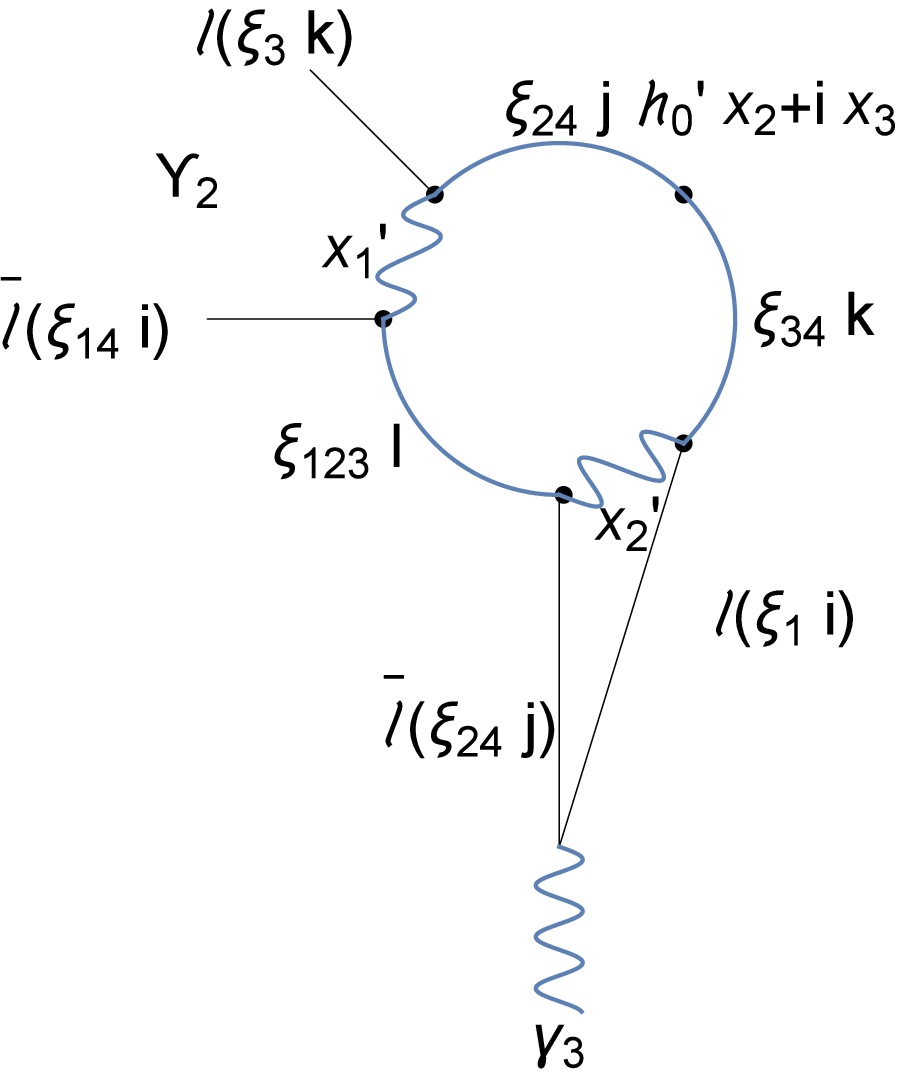}  
\end{center}
\end{minipage}
\begin{minipage}[b]{0.47\linewidth}
\begin{center}
\includegraphics[width=4cm,angle=0,clip]{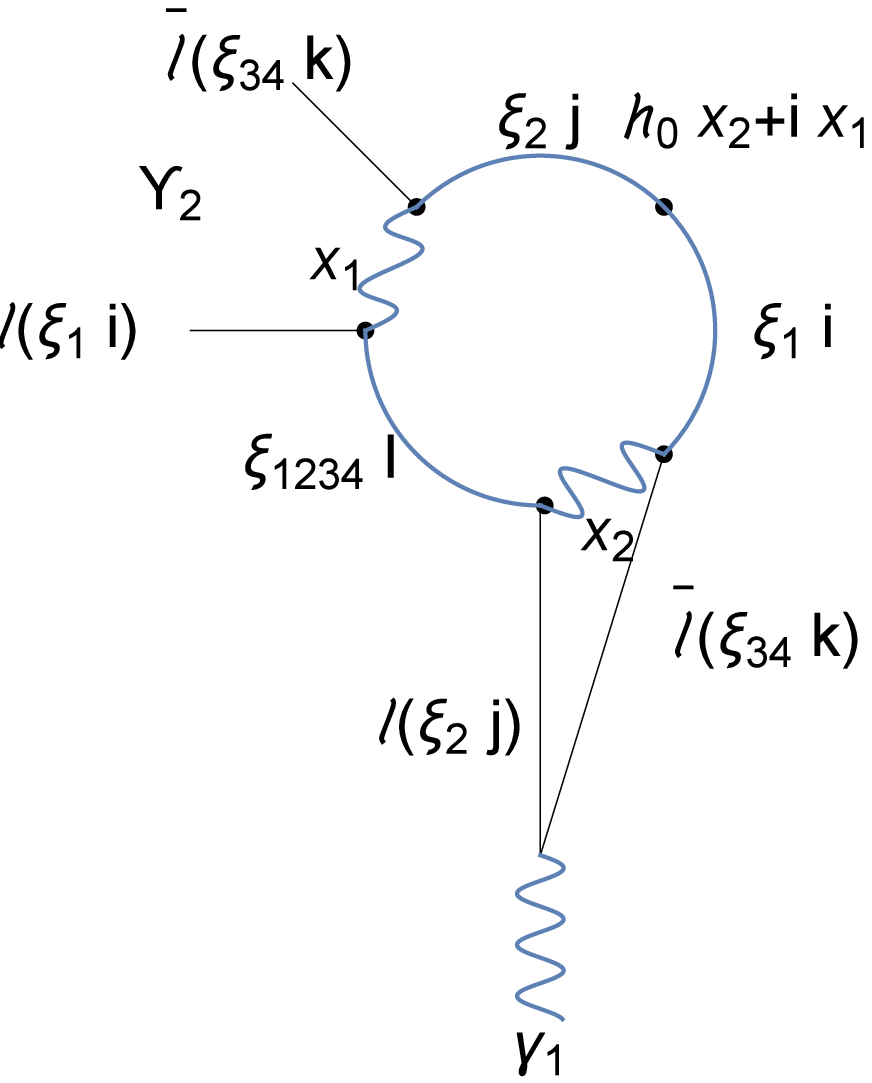}   
\end{center}
\end{minipage}
\hfill
\begin{minipage}[b]{0.47\linewidth}
\begin{center}
\includegraphics[width=4cm,angle=0,clip]{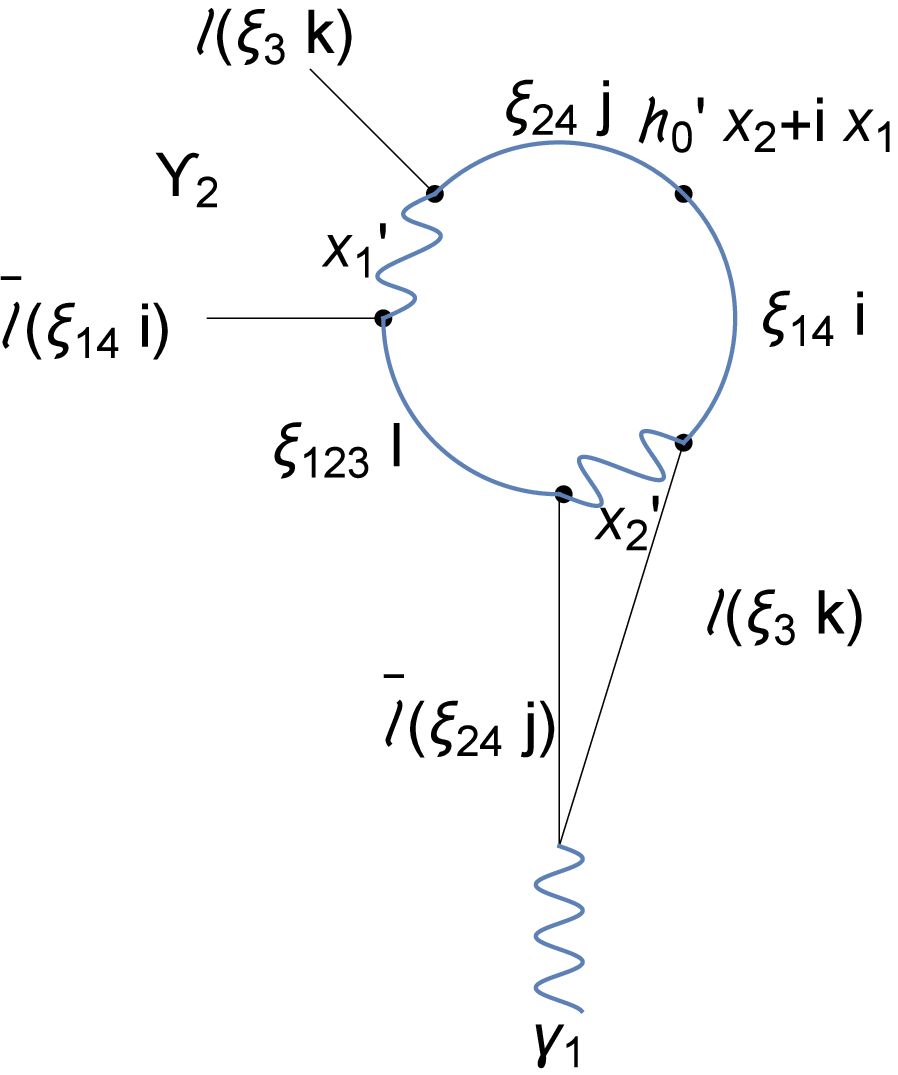}  
\end{center}
\end{minipage}
\caption{The $h(0^+, X+i\,Y)\to \Upsilon_2(k\times i)\gamma_j $ (j=1,3) decay. $\Upsilon_2$ is described by $\ell \bar\ell$ without interactions.  
}
\end{figure}
\begin{figure}[htb]
\begin{minipage}[b]{0.47\linewidth}
\begin{center}
\includegraphics[width=4cm,angle=0,clip]{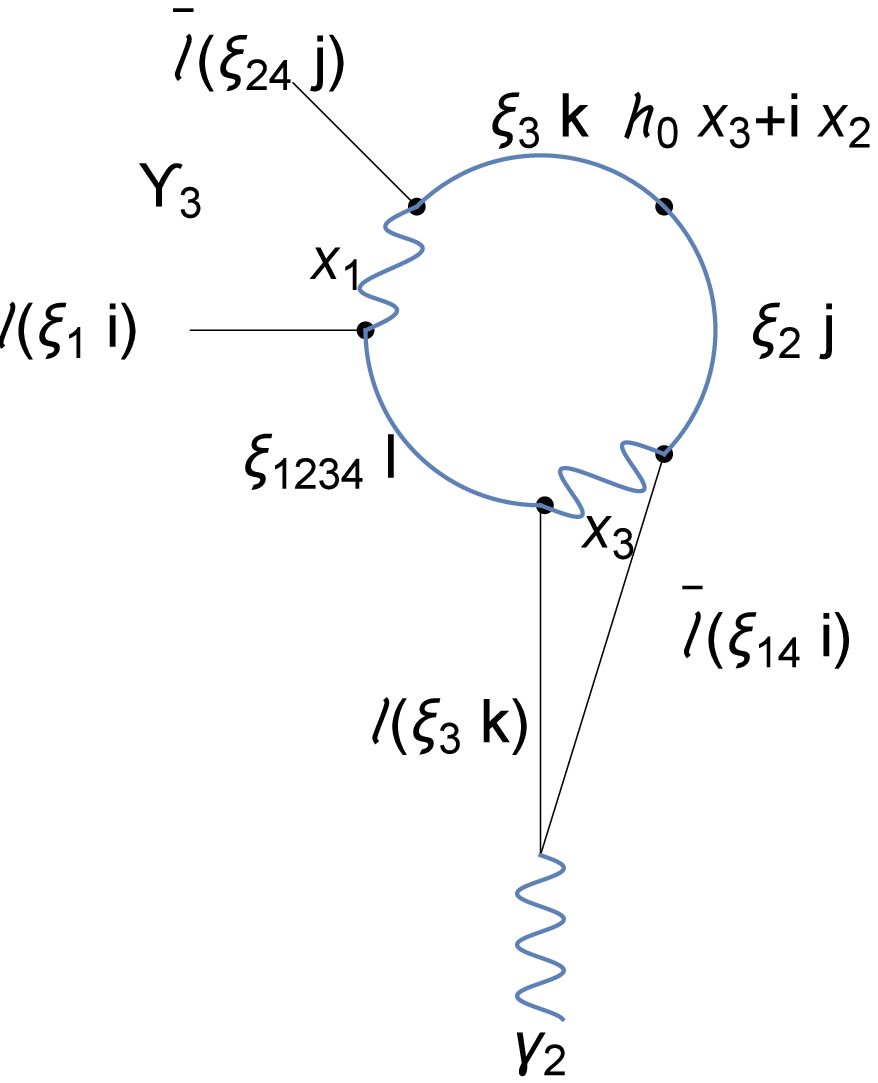}   
\end{center}
\end{minipage}
\hfill
\begin{minipage}[b]{0.47\linewidth}
\begin{center}
\includegraphics[width=4cm,angle=0,clip]{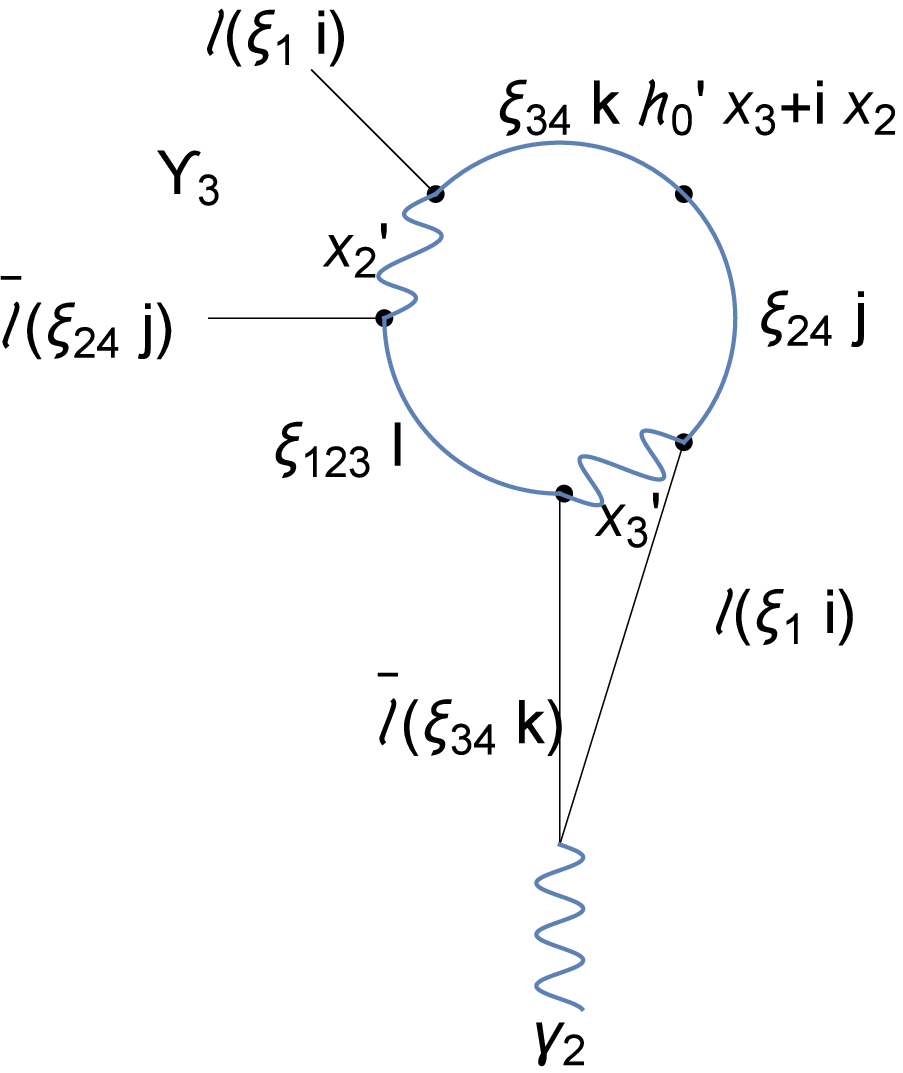}  
\end{center}
\end{minipage}
\begin{minipage}[b]{0.47\linewidth}
\begin{center}
\includegraphics[width=4cm,angle=0,clip]{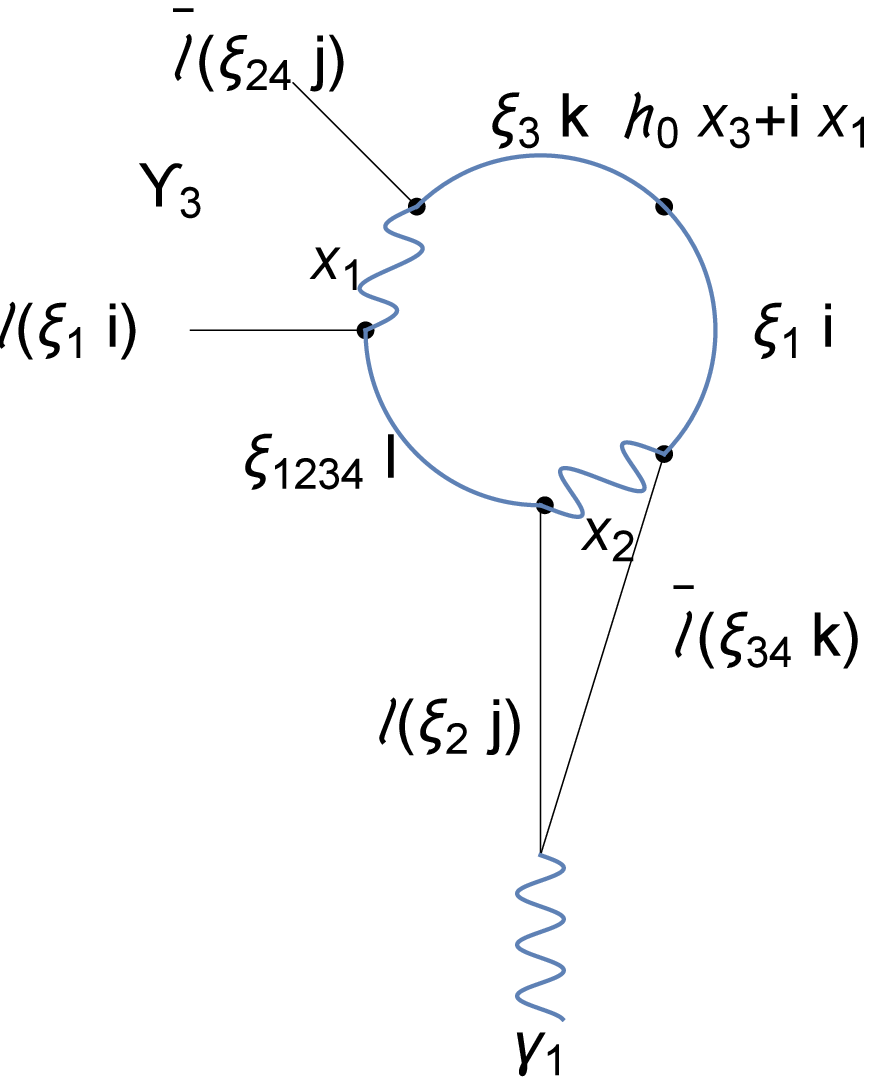}   
\end{center}
\end{minipage}
\hfill
\begin{minipage}[b]{0.47\linewidth}
\begin{center}
\includegraphics[width=4cm,angle=0,clip]{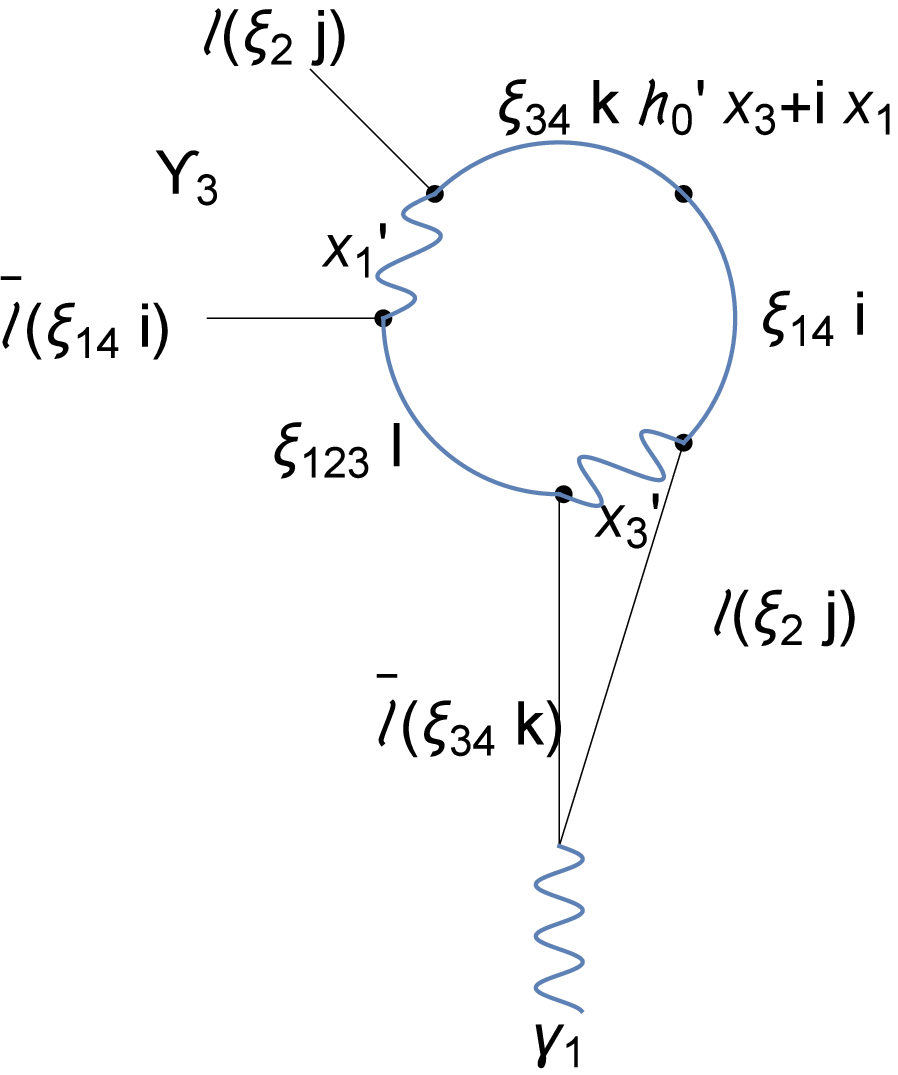}  
\end{center}
\end{minipage}
\label{gb3DD}
\caption{The $h(0^+, X+i\,Y)\to \Upsilon_3(i\times j)\gamma_j $ (j=1,2) decay. $\Upsilon_3$ is described by $\ell\bar\ell$ without interactions.  }
\end{figure}

\section{Discussion and conclusion}
I extended the analysis of $H^0\to\ell\bar\ell\ell\bar\ell\to 2\gamma$ based on Cartan's supersymmetry\cite{SF12a,Cartan66,SF16} to $h^0\to q\bar q q\bar q\to\Upsilon(b\bar b)\gamma(\ell\bar \ell)$.
Cartan's trilinear couplings of quark antiquarks and vector fields predicts apearance of  three polarization states which can correspond to eigen states of  $\Upsilon(1S), \Upsilon(2S)$ and $\Upsilon(3S)$. The $\gamma$ interacts with the $\bar\ell$ produced in the $h^0$ and becomes an $\bar\ell$ and the $\gamma'$ interacts with the $\ell'$ produced in the $h^0$ and becomes a $\ell'$, and the $\bar\ell$ and $\ell'$ pair annihilate and becomes a $\gamma$. 
 
When trilinearity of quark meson field coupling is assumed and the angular momentum of initial $\chi_b$ is $J=0$, smallness of polarizations of $\Upsilon(nS)$ (n=1,2,3) is understandable.  

Coupling of the vector particle and leptons or antileptons is defined by the Lagrangian
\[
\mathcal L=-\frac{1}{4}F_{\mu\nu}F^{\mu\nu}+\sum_{(\ell)=e,\mu,\tau}\bar\psi^{(\ell)}((i\gamma_\mu-ie \,A_\mu)-M^{(\ell)} )\psi^{(\ell)} +h.c.
\]
where
\[
\partial^\mu F_{\mu\nu}(x)=-\sum_{(\ell)=e,\mu,\tau} e \bar\psi^{(\ell)}(x)\gamma_\nu\psi^{(\ell)}(x) +h.c..
\]
We took here $X=A$, and ignored mixing of the vector boson $Z$ and $\gamma$.
 
When the final state $\Upsilon(b\bar b)$ have different helicities, the $\gamma$ produced by $\bar\ell \ell'$ have different helicities, similarly. But there are two different states whether the exchanged vector particle is $x0$ and $b$ and $\bar\ell$ have parallel helicities, or wheteher the exchanged vector particle is $x2$ and $b$ and $\bar\ell$ have different helicities. We expect the $\chi(1P)$ and $\chi(2P)$ assigned in \cite{Abazov12} correspond to the initial states which produces this $\Upsilon(b\bar b)$

In the standard supersymmetric theory, coupling of $t\bar t$ to $b\,\bar W \bar b W$ occurs through $H\bar H$\cite{CMS12}, and the coupling of  $b\bar b, \ell\bar \ell$ through single $h_0$  might seem strange. The standard supersymetric theory cannot reproduce large forward backward asymmetry of $t\bar t$ decay \cite{AAJP15}.  Coupling by single $h_0(x_j+i\,x_k)$ could change the situation. 

Cartan's supersymmetry allows two types of leptons $\psi$ and $\phi$ and their charge conjugates, and the interaction with vector fields is fixed. We extended the model by introducing an effective interaction of  antileptons and quarks or leptons and antiquarks with parallel helicities $x0$, and with different helicities $x2$, in the ${^3P}_0$ states. In the quark pair creation in $p\bar p$ annihilation into mesons, the ${^3P}_0$ model was successful\cite{MFFV87}.  By a detailed comparison of decay modes of $h^0$ and $H^0$, the structure of the Higgs field will become clarified. 

\vskip 0.5 true cm
\leftline{\bf Acknowledgement}

The author thanks Prof. Stan Brodsky for discussions and Dr. Fabian Cruz for sending the informaton of ref.\cite{CMS12}.

\end{document}